# Analytical Model for Metamaterials with Quantum Ingredients

A. Chipouline[1], V. A. Fedotov[2] and A. E. Nikolaenko[2]

[1]*Institute of Applied Physics, Friedrich Schiller University Jena, Max-Wien-Platz 1, D-07743 Jena, Germany*

[2]*Optoelectronics Research Centre and Centre for Photonic Metamaterials, University of Southampton, SO17 1BJ, UK*

**Abstract**

We present an analytical model for describing complex dynamics of a hybrid system consisting of interacting classical and quantum resonant structures. Classical structures in our model correspond to plasmonic nano-resonators of different geometries, as well as other types of nano- and micro-structures optical response of which can be described without invoking quantum-mechanical treatment. Quantum structures are represented by atoms or molecules, or their aggregates (for example, quantum dots and carbon nanotubes), which can be accurately modelled only with the use of quantum approach. Our model is based on the set of equations that combines well-established density matrix formalism appropriate for quantum systems, coupled with harmonic-oscillator equations ideal for modelling sub-wavelength plasmonic and optical resonators. This model can also be straightforwardly adopted for describing electromagnetic dynamics of various hybrid systems outside the photonics realm, such as Josephson-junction metamaterials, or SQUID elements coupled with an RF strip resonator.

## I. Introduction

Accurate description of the dynamics of interacting classical systems is a fundamental task. The current approach is to use a set of coupled equations for two (or more) harmonic oscillators, which can normally be solved under appropriate approximations. It combines mathematical simplicity with adequate physical insight and has been adopted in various branches of science ranging from optics to nuclear physics. In the case of interacting quantum systems their dynamics can be satisfactory described in the framework of quantum mechanics based on the Schrödinger equation or density matrix approach, for instance.

For describing interacting classical and quantum systems a special approach is required. It was previously developed to model the dynamics of lasers where classical system is represented by optical resonator, while quantum system – by amplifying medium. The basic idea here is that the quantum formalism allows accurate calculation of the medium's polarizability, which can than be used in the Maxwell equations describing electromagnetic fields in the optical resonator [1].

With the development of nanotechnology it has become possible to engineer and study quantum-classical systems at the nanoscale, such as metallic nano-resonators, which are still classical, hybridised with quantum dots, carbon nanotubes or dye molecules [2 - 5]. Although optical response of the metallic nano-resonator is affected by plasmonic excitations and shape effects its rather complicated dynamics can still be satisfactory modelled by the harmonic oscillator equations with appropriately chosen parameters [6]. This allows to extend quantum-classical description for the case of individual plasmonic nano-resonators and metamaterials (i.e. assemblies of such resonators) combined with various quantum systems and build a model, which can describe a wide range of optical and plasmonic effects resulting from the hybridisation such as, for example, loss compensation, enhancement of nonlinear response and luminescence in metamaterials, etc. Furthermore, the dynamics of superconducting Josephson-junction-based metamaterials, as well as SQUIDs coupled to an RF strip resonator [7] can be described using the same model.

**II. Model**

In this paper we consider a Quantum System (QS) placed in the near-field zone of a Classical electromagnetic System (CS). The field produced by the CS, $E_{CS}$, affects QS that in turn acts on CS with its field $E_{QS}$. In addition, there is an external field $E_{ext}$ of the incident light, which interacts with both CS and QS.

The actual number of the harmonic-oscillator equations required for the adequate modelling CS depends on the particular structure of CS [6]. For the illustration purpose we will restrict our analysis to just one harmonic-oscillator equation, which should not limit the generality of our model however. The dynamics of QS is modelled using density matrix formalism.

In general case the quantum dynamics of QS that is assumed to be in contact with a thermostat is described by the following set of ordinary differential equations [8]:

$$\begin{cases} \dfrac{d\rho_{nn}}{dt} + \sum_m \left( k_{nm} \rho_{nn} - k_{mn} \rho_{mm} \right) = -\dfrac{i}{\hbar} \sum_m \left( H_{nm} \rho_{mn} - H_{mn} \rho_{nm} \right) \\ \dfrac{d\rho_{kl}}{dt} + i\omega_{kl} \rho_{kl} + \dfrac{\rho_{kl}}{\tau_{kl}} = -\dfrac{i}{\hbar} \sum_m \left( H_{km} \rho_{ml} - H_{ml} \rho_{km} \right) \end{cases} \quad (1)$$

Here $k_{nm}$ and $\tau_{kl}$ are energy and phase relaxation constants respectively, $\omega_{kl}$ is frequency of the transition from $k$ to $l$, $\rho_{nn}$ and $\rho_{kl}$ ($\rho_{kl} = \rho^{*}_{kl}$) are diagonal and non-diagonal elements of the density matrix, $H_{kl}$ is a Hamiltonian matrix element responsible for interaction of the quantum system with the external field.

In framework of this formalism an averaged polarisation density is expressed through non-diagonal density matrix elements:

$$P_{kl} = N \mu_{kl} \left( \rho_{kl} + \rho_{lk} \right) \quad (2)$$

$\mu_{kl}$ is the dipole moment of a quantum system, which is proportional to overlap integral between psi-functions of both levels, $N$ is the quantum systems concentration.

In the case of resonant interaction the QS internal dynamics can be modelled to a first approximation by a two-level system subjected to a pump:

$$\begin{cases} \dfrac{d\rho_{12}}{dt} - i\omega_{21}\rho_{12} + \dfrac{\rho_{12}}{\tau_2} = -\dfrac{iH_{12}(\rho_{22}-\rho_{11})}{\hbar} \\ \dfrac{d\rho_{22}}{dt} + \dfrac{\rho_{22}}{\tilde{\tau}_1} = -\dfrac{iH_{12}(\rho_{12}-\rho_{12}^*)}{\hbar} + W\rho_{11} \\ \rho_{11} + \rho_{22} = 1 \end{cases} \quad (3)$$

Here $\rho_{22}$, $\rho_{11}$ and $\rho_{12}$, $\rho_{12}^*$ are the diagonal and non-diagonal matrix density elements, respectively; $\tau_2$ and $\tau_1$ are the constants describing phase and energy relaxation processes due to the interaction with a thermostat; $\omega_{21} = (E_2 - E_1)/\hbar$ is the transition frequency between levels 2 and 1; $H_{12}$ is the Hamiltonian matrix element responsible for interaction of QS with the external fields; $W$ is the phenomenological pump rate – this could model pumping QS. It is also convenient to introduce new variables $N = \rho_{22} - \rho_{11}$ and $N_0 = \dfrac{W\tau_1 - 1}{W\tau_1 + 1}$ :

$$\begin{cases} \dfrac{d\rho_{12}}{dt} - i\omega_{21}\rho_{12} + \dfrac{\rho_{12}}{\tau_2} = -\dfrac{iH_{12}N}{\hbar} \\ \dfrac{dN}{dt} + \dfrac{N-N_0}{\tau_1} = -\dfrac{2iH_{12}(\rho_{12}-\rho_{21})}{\hbar} \\ \tau_1 = \dfrac{\tilde{\tau}_1}{W\tilde{\tau}_1 + 1} \end{cases} \quad (4)$$

In order to describe dynamics in the plasmonic nano-resonator we use the following harmonic-oscillator equation:

$$\dfrac{d^2x}{dt^2} + 2\gamma\dfrac{dx}{dt} + \omega_0^2 x = \chi(E_{ext} + E_{QS}) \quad (5)$$

Here $\gamma$ and $\omega_0$ are the loss coefficient and resonance eigen-frequency, $E_{ext}$ and $E_{QS}$ are the external electric field and field generated by QS respectively, and $\chi^{-1}$ is the effective kinetic inductance of the nano-resonator.
From Eqs (4) and (5) we obtain:

$$\begin{cases} \dfrac{d\rho_{12}}{dt} - i\omega_{21}\rho_{12} + \dfrac{\rho_{12}}{\tau_2} = -\dfrac{iH_{12}N}{\hbar} \\ \dfrac{dN}{dt} + \dfrac{N-N_0}{\tau_1} = -\dfrac{2iH_{12}(\rho_{12}-\rho_{21})}{\hbar} \\ \dfrac{d^2x}{dt^2} + 2\gamma\dfrac{dx}{dt} + \omega_0^2 x = \chi\left(E_{ext} + E_{QS}\right) \end{cases} \quad (6)$$

In order to make a next step it is necessary to determine the nature of the interaction between CS and QS and write down expressions for $H_{12}$ and $E_{QS}$. We assume that the fields of both systems are predominantly electric in the near-field zone and produced by the effective electric dipole moments **d**. Electric field of an oscillating dipole is inversely proportional to the third power of the distance $R$ between CS and QS, namely:

$$E = \frac{d}{R^3} \quad (7)$$

Correspondingly the dipole moment of QS generates electric field, which can be written as:

$$E_{QS} = \frac{d_{QS}}{R^3} = \frac{\mu_{QS}(\rho_{12}+\rho_{21})}{R^3}, \quad (8)$$

where $\mu_{QS}$ is the dipole moment of QS.
Electric field of CS according to the same relation is:

$$E_{CS} = \frac{d_{CS}}{R^3} = \frac{\mu_{CS} x}{R^3}, \quad (9)$$

where $\mu_{CS}$ is the dipole moment of CS. The dimensionless variable $x$ has the same physical meaning here as non-diagonal element of the density matrix, namely dimensionless polarization, which becomes clear from comparing Eqs. (9) and (8).
The Hamiltonian of interaction $H_{12}$ is defined by the following expressions:

$$\begin{cases} H_{12} = -\mu_{QS}\left(E_{ext} + E_{CS}\right) = -\left(\mu_{QS} E_{ext} + \alpha_x x\right) \\ \alpha_x = \dfrac{\mu_{QS}\mu_{CS}}{R^3} \end{cases} \quad (10)$$

Substituting Eqs. (8) and (10) into (6) we obtain:

$$\begin{cases} \dfrac{d\rho_{12}}{dt} - i\omega_{21}\rho_{12} + \dfrac{\rho_{12}}{\tau_2} = \dfrac{i(\mu_{QS} E_{ext} + \alpha_x x)N}{\hbar} \\ \dfrac{dN}{dt} + \dfrac{N - N_0}{\tau_1} = \dfrac{2i(\mu_{QS} E_{ext} + \alpha_x x)(\rho_{12} - \rho_{21})}{\hbar} \\ \dfrac{d^2 x}{dt^2} + 2\gamma \dfrac{dx}{dt} + \omega_0^2 x - \alpha_\rho (\rho_{12} + \rho_{21}) - \chi E_{ext} = 0 \\ N_0 = \dfrac{(W\tilde{\tau}_1 - 1)}{(W\tilde{\tau}_1 + 1)}, \quad \tau_1 = \dfrac{\tilde{\tau}_1}{W\tilde{\tau}_1 + 1} \\ \alpha_\rho = \dfrac{\mu_{QS} \chi}{R^3} \\ \alpha_x = \dfrac{\mu_{QS} \mu_{CS}}{R^3} \end{cases} \quad (11)$$

Here $N_0$ is the population inversion due to pump (in the absence of pump $N_0 = -1$); $N_0 > 0$ corresponds to the regime of amplification, $N_0 < 0$ - to losses. Both eigen-frequencies $\omega_{21}$ and $\omega_0$ are the resonance frequencies of QS and CS respectively and can vary independently. Rotating wave approximation for the system (11) is introduced through the following notations:

$$\begin{cases} \rho_{12} = \tilde{\rho}_{12} \exp(i\omega t) \\ x = \dfrac{1}{2}\left(\tilde{x}(t)\exp(-i\omega t) + \tilde{x}(t)^* \exp(i\omega t)\right) \\ E_{ext} = \dfrac{1}{2}\left(A(t)\exp(-i\omega t) + A(t)^* \exp(i\omega t)\right) \end{cases} \quad (12)$$

resulting in:

$$\begin{cases} \dfrac{d\tilde{\rho}_{12}}{dt} + \tilde{\rho}_{12}\left(\dfrac{1}{\tau_2} + i(\omega - \omega_{21})\right) = \dfrac{i\alpha_x \tilde{x}^* N}{\hbar} + \dfrac{i\mu_{QS} A^* N}{\hbar} \\ \dfrac{dN}{dt} + \dfrac{(N - N_0)}{\tau_1} = \dfrac{i\alpha_x (\tilde{x}\tilde{\rho}_{12} - \tilde{x}^* \tilde{\rho}_{12}^*) + i\mu_{QS}(A\tilde{\rho}_{12} - A^* \tilde{\rho}_{12}^*)}{2\hbar} \\ 2(\gamma - i\omega)\dfrac{d\tilde{x}}{dt} + (\omega_0^2 - \omega^2 - 2i\omega\gamma)\tilde{x} = \alpha_\rho \tilde{\rho}_{12}^* + \chi A \end{cases} \quad (13)$$

Eqs. (13) is the master set of equations describing regular dynamics of interacting QS and CS. Taking into account stochastic noise sources, the set becomes:

$$\begin{cases} \dfrac{d\tilde{\rho}_{12}}{dt} + \tilde{\rho}_{12}\left(\dfrac{1}{\tau_2} + i(\omega - \omega_{21})\right) = \dfrac{i\alpha_x \tilde{x}^* N}{\hbar} + \dfrac{i\mu_{QS} A^* N}{\hbar} + \xi_\rho \\ \dfrac{dN}{dt} + \dfrac{(N - N_0)}{\tau_1} = \dfrac{i\alpha_x\left(\tilde{x}\tilde{\rho}_{12} - \tilde{x}^*\tilde{\rho}_{12}^*\right) + i\mu_{QS}\left(A\tilde{\rho}_{12} - A^*\tilde{\rho}_{12}^*\right)}{2\hbar} \\ 2(\gamma - i\omega)\dfrac{d\tilde{x}}{dt} + (\omega_0^2 - \omega^2 - 2i\omega\gamma)\tilde{x} = \alpha_\rho \tilde{\rho}_{12}^* + \chi A + \xi_x \end{cases} \quad (14)$$

Here $\xi_\rho$ and $\xi x$ are the stochastic Langevin terms, which take into account spontaneous emission and thermal fluctuations respectively.

## III. Applications of the model

The set of equations (14) can describe the following experimental situations:

1. **Nano-laser (spaser) [2, 9 - 11].** In this case $N_0 = \dfrac{W\tau_1 - 1}{W\tau_1 + 1} > 0$ and $A=0$, Eq. (14) gives transition and stationary dynamics of a nano-laser (spaser):

$$\begin{cases} \dfrac{d\tilde{\rho}_{12}}{dt} + \tilde{\rho}_{12}\left(\dfrac{1}{\tau_2} + i(\omega - \omega_{21})\right) = \dfrac{i\alpha_x \tilde{x}^* N}{\hbar} + \xi_\rho \\ \dfrac{dN}{dt} + \dfrac{(N - N_0)}{\tau_1} = \dfrac{i\alpha_x\left(\tilde{x}\tilde{\rho}_{12} - \tilde{x}^*\tilde{\rho}_{12}^*\right)}{2\hbar} \\ 2(\gamma - i\omega)\dfrac{d\tilde{x}}{dt} + (\omega_0^2 - \omega^2 - 2i\omega\gamma)\tilde{x} = \alpha_\rho \tilde{\rho}_{12}^* + \xi_x \end{cases} \quad (15)$$

With the stochastic Langevin terms one can calculate laser bandwidth in analog with well-known Schawlow-Towns approach [12].

2. **Luminescence enhancement [4, 13].** As in the case of spaser $N_0 = \dfrac{W\tau_1 - 1}{W\tau_1 + 1} > 0$, but $A \neq 0$ is assumed to be small (no saturation) describing zero field fluctuations causing spontaneous emission. The Purcell effect appears as a multiplicator for the field $A$ (here denoted by $F$) and takes into account density of states detuning for this field. It has to be emphasized that the Purcell effect does not affect both relaxation times $\tau_2$ and $\tau_1$, which appear from the interaction with thermostat (until the density of states of the thermostat is changed by the nano-resonator):

$$\begin{cases} \dfrac{d\tilde{\rho}_{12}}{dt} + \tilde{\rho}_{12}\left(\dfrac{1}{\tau_2} + i(\omega - \omega_{21})\right) = \dfrac{i\alpha_x \tilde{x}^* N_0}{\hbar} + \dfrac{i\mu_{QS} A^* N_0}{\hbar} F + \xi_\rho \\ 2(\gamma - i\omega)\dfrac{d\tilde{x}}{dt} + (\omega_0^2 - \omega^2 - 2i\omega\gamma)\tilde{x} = \alpha_\rho \tilde{\rho}_{12}^* + \chi A F + \xi_x \end{cases} \quad (16)$$

3. **Nonlinear response enhancement [5, 9].** The nonlinearity of QS appears due to the saturation effect and basically does not require either positive $N_0$ or nano-resonator. Enhancement of the saturation is caused by an addition channel: external field transfers energy to QS through the nano-resonator in addition to the direct pumping. Taking into account the field enhancement effect near the plasmonic nano-resonator, the model adequately describes the increased strength of the nonlinear response experimentally observed in [5, 9]:

$$\begin{cases} \dfrac{d\tilde{\rho}_{12}}{dt} + \tilde{\rho}_{12}\left(\dfrac{1}{\tau_2} + i(\omega - \omega_{21})\right) = \dfrac{i\alpha_x \tilde{x}^* N}{\hbar} + \dfrac{i\mu_{QS} A^* N}{\hbar} \\ \dfrac{dN}{dt} + \dfrac{(N - N_0)}{\tau_1} = \dfrac{i\alpha_x \left(\tilde{x}\tilde{\rho}_{12} - \tilde{x}^*\tilde{\rho}_{12}^*\right) + i\mu_{QS}\left(A\tilde{\rho}_{12} - A^*\tilde{\rho}_{12}^*\right)}{2\hbar} \\ 2(\gamma - i\omega)\dfrac{d\tilde{x}}{dt} + (\omega_0^2 - \omega^2 - 2i\omega\gamma)\tilde{x} = \alpha_\rho \tilde{\rho}_{12}^* + \chi A \end{cases} \quad (17)$$

4. **Enhancement of magnetic dipolar response [14].** Marginal modification of system (13) allows to model the enhancement of high-order multipole response in the hybrid metamaterial. In particular, complex nano-resonators (like double-wire or split-ring resonators) support anti-symmetric mode of excitation, which is responsible for magnetic dipolar response [6]. It can be adequately described by two (instead of one) coupled harmonic oscillator equations. In the case of sufficiently strong pumping $N_0 = \dfrac{W\tau_1 - 1}{W\tau_1 + 1} > 0$ the energy transferred from the appropriately positioned QS will support excitation of the anti-symmetric mode.

5. **Quantum magnetic metamaterials [14].** Combining active QS (such as quantum dots) with the specially designed plasmonic nano-resonators can lead to magnetization at optical frequencies (see also **"Enhancement of magnetic dipolar response"** above) produced not only by the plasmonic modes, but also modes of coherently coupled QS. Such hybrid structures could serve as building blocks for the lossless metamaterials with strong magnetic response at optical frequencies.

6. **Linear and nonlinear response of SQUIDs [7, 15].** The behavior of SQUID coupled with an RF resonator [15] also falls in the range of phenomena described by the model. In this particular case the dynamics of SQUID is governed by the direct interaction with the resonator without influence from the external field:

$$\begin{cases} \dfrac{d\tilde{\rho}_{12}}{dt} + \tilde{\rho}_{12}\left(\dfrac{1}{\tau_2} + i(\omega - \omega_{21})\right) = \dfrac{i\alpha_x \tilde{x}^* N}{\hbar} \\ \dfrac{dN}{dt} + \dfrac{(N - N_0)}{\tau_1} = \dfrac{i\alpha_x\left(\tilde{x}\tilde{\rho}_{12} - \tilde{x}^*\tilde{\rho}_{12}^*\right) + i\mu_{QS}\left(A\tilde{\rho}_{12} - A^*\tilde{\rho}_{12}^*\right)}{2\hbar} \\ 2(\gamma - i\omega)\dfrac{d\tilde{x}}{dt} + \left(\omega_0^2 - \omega^2 - 2i\omega\gamma\right)\tilde{x} = \alpha_\rho \tilde{\rho}_{12}^* + \chi A \end{cases} \quad (18)$$